\documentstyle[twoside,12pt]{article}

\catcode`\@=11
\long\def\@makefntext#1{ %\parindent 1em
\protect\noindent \hbox to 3.2pt {\hskip-.9pt
$^{{\eightrm\@thefnmark}}$\hfil}#1\hfill} %can be used

 \def\@makefnmark{\hbox to 0pt{$^{\@thefnmark}$\hss}}  %original

\def\ps@myheadings{\let\@mkboth\@gobbletwo
\def\@oddhead{\hbox{} %\sl
\rightmark\hfil\eightrm\thepage}
\def\@oddfoot{}\def\@evenhead{\eightrm\thepage\hfil %\sl
\leftmark\hbox{}}\def\@evenfoot{}
\def\sectionmark##1{}\def\subsectionmark##1{}}

\oddsidemargin=\evensidemargin
\newcounter{sectionc}\newcounter{subsectionc}\newcounter{subsubsectionc}
\renewcommand{\section}[1] {\vspace{12pt}\addtocounter{sectionc}{1}
\setcounter{subsectionc}{0}\setcounter{subsubsectionc}{0}\noindent
	{\bf\thesectionc. #1}\par\vspace{5pt}}
\renewcommand{\subsection}[1] {\vspace{12pt}\addtocounter{subsectionc}{1}
	\setcounter{subsubsectionc}{0}\noindent
	{\bf\thesectionc.\thesubsectionc. {\kern1pt \bf\it #1}}\par\vspace{5pt}}
\renewcommand{\subsubsection}[1] {\vspace{12pt}\addtocounter{subsubsectionc}{1}
	\noindent{\thesectionc.\thesubsectionc.\thesubsubsectionc.
	{\kern1pt \it #1}}\par\vspace{5pt}}
\newcommand{\nonumsection}[1] {\vspace{12pt}\noindent{\bf #1}
	\par\vspace{5pt}}

\newcommand{\textlineskip}{\baselineskip=14pt}
\newcommand{\smalllineskip}{\baselineskip=12pt}

\def\eightcirc{
\begin{picture}(0,0)
\put(4.4,1.8){\circle{6.5}}
\end{picture}}
\def\eightcopyright{\eightcirc\kern2.7pt\hbox{\eightrm c}}

\newcounter{itemlistc}
\newcounter{romanlistc}
\newcounter{alphlistc}
\newcounter{arabiclistc}

\newcommand{\fcaption}[1]{
        \addtocounter{figure}{1}
         {{\tenrm Fig.~\thefigure . #1} }\hfil\break }

\newcommand{\tcaption}[1]{			%CENTRALISE TABLE CAPTION
        \addtocounter{table}{1}
         {{\tenrm\offinterlineskip Table~\thetable . #1} }\hfil\break }

\def\thebibliography#1{\nonumsection{\large \bf References}\list
  {[\arabic{enumi}]}{\settowidth\labelwidth{[#1]}\leftmargin\labelwidth
    \advance\leftmargin\labelsep
    \usecounter{enumi}}
    \def\newblock{\hskip .11em plus .33em minus .07em}
    \sloppy\clubpenalty4000\widowpenalty4000}

\def\pmb#1{\setbox0=\hbox{#1}
	\kern-.025em\copy0\kern-\wd0
	\kern.05em\copy0\kern-\wd0
	\kern-.025em\raise.0433em\box0}

\def\fnt#1#2{\footnotetext{\kern-.3em
	{$^{\mbox{\scriptsize #1}}$}{#2}}}

\def\fpage#1{\begingroup
\voffset=.3in
\thispagestyle{empty}\begin{table}[b]\centerline{\footnotesize #1}
	\end{table}\endgroup}

 1
 1
 1

\font\eightrm=cmr8

\def\qed{\hbox{${\vcenter{\vbox{                          %HOLLOW SQUARE
   \hrule height 0.4pt\hbox{\vrule width 0.4pt height 6pt
   \kern5pt\vrule width 0.4pt}\hrule height 0.4pt}}}$}}

\newcommand{\be}{\begin{eqnarray}}
\newcommand{\ee}{\end{eqnarray}}
\newcommand{\dslash}{\partial \hskip -0.5em /}
\newcommand{\Dslash}{D \hskip -0.7em /}

\newcommand{\tr}{{\rm tr}}
\newcommand{\Tr}{{\rm Tr}}

\newcommand{\A}{{\cal A}}

\input psfig

\begin{document}
\normalsize\textlineskip
{\thispagestyle{empty}
\setcounter{page}{1}

\fpage{1}
\rightline{UNITU-THEP-9/1993}
\rightline{hep-ph/9308327}
\rightline{August 1993}
\vspace{1cm}
\centerline{\LARGE  $1/N_C$ Corrections to $g_A$ in the
Light of PCAC$^\dagger $}
\vspace{0.9cm}
\centerline{R. Alkofer and H. Weigel}
\vspace{0.9cm}
\centerline{Institute for Theoretical Physics}
\vspace{0.3cm}
\centerline{T\"ubingen University}
\vspace{0.3cm}
\centerline{Auf der Morgenstelle 14}
\vspace{0.3cm}
\centerline{D-72076 T\"ubingen, FR Germany}

\vspace{4cm}
\normalsize\textlineskip
\noindent
\centerline{\bf Abstract}
\vspace{0.5cm}

\noindent
We comment on recently discovered
$1/N_C$ corrections to the nucleon axial current in
the framework of the Nambu--Jona-Lasinio soliton. This kind of corrections
arises only for a special treatment of ordering ambiguities of collective
coordinates and operators in the semiclassical quantization.
In addition to the missing derivation of this special quantization
scheme from first principles its na\"\i ve application violates
partial conservation of the axial current (PCAC). We show how
within this scheme PCAC can be restored and determine the
corresponding $1/N_C$ corrections to the equation of motion
for the chiral soliton. The
resulting self-consistent solution allows to evaluate the nucleon
axial coupling constant $g_A$ directly as a matrix element of the
axial current as well as indirectly from the pion profile function.
Enhancement is found for those baryon properties which are sensitive
to the long range behavior of the pion profile.

\vfill

\noindent
$^\dagger $
{\footnotesize{Supported by the Deutsche Forschungsgemeinschaft (DFG) under
contract Re 856/2-1.}}
\eject

Chiral symmetry and its spontaneous breaking are generally accepted as
guiding principles for the construction of models describing hadron
physics. A prominent feature of chiral symmetry is the partial conservation
of the axial vector current (PCAC). PCAC relates the divergence of the axial
vector current $A_\mu^a$ to the pseudoscalar-isovector pion field $\pi^a$:
\begin{eqnarray}
\partial^\mu A_\mu^a = m_\pi^2 f_\pi \pi^a.
\label{pcac1}
\end{eqnarray}
Here $m_\pi=138$MeV and $f_\pi=93$MeV denote the pion mass and decay
constant, respectively. Since the axial current is of the nature of
a Noether current its (partial) conservation heavily relies on the
pion fields satisfying their equation of motion; {\it i.e.}
eqn.\ (\ref{pcac1}) is equivalent to the equation of motion for the pion
field. Therefore variations of the axial current imply corresponding
changes in the equation of motion for the pion field.

In baryon physics the axial current plays an important role since
its matrix element between nucleon states, the nucleon axial
coupling constant $g_A$, directly  relates to the neutron $\beta$
decay amplitude. It has been a long standing problem for chiral
soliton models of baryons that the predicted value for $g_A$ is
only of the order 0.6 to 0.9 underestimating the experimental
value 1.26\cite{ho93}\footnote{Note, however, that in the chiral
quark model $g_A$ is somewhat overestimated \cite{ja88}.}.
Recently it has been claimed \cite{wa93} that $1/N_C$ corrections
might provide the missing amount to match the empirical value.
However, these corrections are only present if a certain ordering
prescription for the generators appearing in the semiclassical
quantization scheme is adopted. We will resolve this point below.
Additionally, the way these corrections are treated in ref.\
\cite{wa93} violates PCAC.
In that calculation only the corrections of the
axial current were taken into account disregarding the
corresponding changes of the soliton profile. In this letter we
will demonstrate that these corrections are mitigated but nevertheless
sizable when PCAC is properly accounted for.

Let us briefly review the connection between the equation of motion
for the meson profile function and PCAC in chiral soliton models. The
static solution to the Euler-Lagrange equations acquires the
hedgehog shape for the non-linear representation of the meson fields:
\begin{eqnarray}
U_0(\mbox {\boldmath $r $}) = {\rm exp}
\left(i \mbox{\boldmath $\tau$} \cdot{\bf \hat r}\ \Theta(r)\right),
\label{hedgehog}
\end{eqnarray}
$\Theta(r)$ being the chiral angle. The equation of motion
for $\Theta(r)$ may generally be expressed in the
form ${\cal F}(\Theta)=0$. The hedgehog configuration
(\ref{hedgehog}) is quantized by introducing collective coordinates
for the (iso-) rotational zero modes:
\begin{eqnarray}
U(\mbox {\boldmath $r $},t) =
R(t) U_0(\mbox {\boldmath $r $})R^\dagger(t), \qquad
R(t)\in SU(2).
\label{colcor}
\end{eqnarray}
Due to the vanishing time component of the axial current for the
configuration (\ref{colcor}) PCAC reduces to\footnote{For our
sign conventions the pion field of the hedgehog configuration
is given by $\pi^a = - f_\pi D_{ab} \hat r_b {\rm sin} \Theta$.}
\begin{eqnarray}
\partial_i A_i^a = D_{ab} \hat r_b {\cal F}(\Theta)
%\times \left({\rm equation\ of\ motion\ for}\ \Theta\right)
- f_\pi^2 m_\pi^2 D_{ab}\hat r_b {\rm sin} \Theta.
\label{pcac2}
\end{eqnarray}
$D_{ab}=(1/2){\rm tr}({\mbox{\boldmath $\tau$}}_a R
{\mbox{\boldmath $\tau$}}_b R^\dagger)$ denotes the
adjoint representation of the collective rotation.
The PCAC relation is obviously recovered if the chiral angle
satisfies the appropriate equation of motion which is
obtained upon inversion of eqn. (\ref{pcac2}):
\begin{eqnarray}
0={\cal F}(\Theta)=\hat r_b D_{ab}\partial_i A_i^a
+ f_\pi^2 m_\pi^2 {\rm sin} \Theta.
\label{eqm}
\end{eqnarray}

The axial coupling constant $g_A$ is defined as the matrix element of
the axial current between proton states at zero momentum transfer:
\begin{eqnarray}
g_A=2 \langle  p\uparrow |A_3^3|
p\uparrow\rangle .
\label{defga}
\end{eqnarray}
The defining equation for the pion decay constant provides the
asymptotic form of the axial current:
\begin{eqnarray}
\lim_{r\to \infty} A_i^a = f_\pi \lim_{r\to \infty} \partial_i \pi^a
= - f_\pi^2 \lim_{r\to \infty} \partial_i \left(D_{ab} \hat r_b
\Theta \right).
\label{asympt}
\end{eqnarray}
The identity $\partial_j(r_i A_j^a)=A_i^a+r_j\partial_i A_i^a$ then
allows to determine $g_A$ from the pion profile via PCAC (\ref{pcac2}):
\begin{eqnarray}
g_A =  \frac{4\pi}{3} f_\pi^2\lim_{R\to\infty}R^3
\frac{\partial \Theta}{\partial r}\Big|_{r=R}
-\frac{8\pi}{9} f_\pi^2 m_\pi^2\int dr r^3 {\rm sin}\Theta
\label{tail}
\end{eqnarray}
wherein extensive use has been made of the matrix element
$\langle D_{33} \rangle_{\rm proton}=-1/3$ in the two flavor case. Note
the appearance of the factor 3/2 in the surface term which arises at
zero momentum transfer in the chiral limit ($m_\pi=0$) \cite{ad83}. We
would like to emphasize that the equivalence of eqns. (\ref{defga}) and
(\ref{tail}) manifests a direct consequence of PCAC for chiral soliton models.
Phrased otherwise: PCAC allows to read off $g_A$ directly and uniquely
from the pion
profile.

So far these considerations have been fairly general and apply to
various kinds of chiral soliton models. Let us now turn to the specific
case of a two flavor Nambu--Jona-Lasinio (NJL)
 model of scalar and pseudoscalar interactions in the isospin limit
\cite{na61}.
After bosonization \cite{eb86} the action ${\cal A}_{NJL}$ may be expressed
as a sum $\A_{NJL}=\A_F+\A_m$ of a fermion determinant\footnote{For
simplicity we treat only the chiral field $U$ as space--dependent quantity
and keep the scalar field equal to its vacuum expectation value
$\langle \phi \rangle =M$.}
\be
\A_F=\Tr\log_\Lambda(i\Dslash)
=\Tr\log_\Lambda\left(i\dslash-MU^{\gamma_5}\right)
\label{fdet}
\ee
and a purely mesonic part
\be
\A_m=\frac{mM}{4G}\int d^4x \tr \left(U+U^{\dag}-2 \right)
\label{ames}
\ee
where $G$ is the NJL coupling constant. For a finite pion mass it may be
eliminated via the relation $G=mM/m_\pi^2f_\pi^2$. For a given constituent
quark mass $M$ the cut-off $\Lambda$ is determined such as to reproduce
the empirical value of the pion decay constant while the current quark
mass $m$ is obtained by solving the gap equation \cite{eb86}. Throughout
this letter we will apply Schwinger's proper time regularization
prescription \cite{sch51} to the diverging fermion determinant. The NJL
model supports static solitons (\ref{hedgehog}) which are obtained as
the solutions of the Euler-Lagrange equations to the static energy functional
\cite{re88,me89,al90}.

Only the fermion determinant (\ref{fdet}) contributes to the axial current.
Since ${\cal A}_F={\cal A}_{\rm val}+{\cal A}_{\rm sea}$ splits into
valence and sea quark parts we have a corresponding decomposition of the
axial current:
\begin{eqnarray}
A_i^a = A_i^{{\rm (val)}a} + A_i^{{\rm (sea)}a}.
\label{valsea}
\end{eqnarray}
The valence quark part is given by
\begin{eqnarray}
A_i^{{\rm (val)}a} = N_C {\bar \Psi}_{\rm (val)}\gamma_i\gamma_5
\frac{\tau^a}{2}\Psi_{\rm (val)}
\label{aval1}
\end{eqnarray}
with $N_C$ being the number of color degrees of freedom.
In order to determine the valence quark wave function $\Psi_{\rm (val)}$
we need to construct the eigenfunctions $\psi_\mu$ of the static
Hamiltonian $h_0$:
\begin{eqnarray}
h_0 \psi_\mu =
\left(\mbox {\boldmath $\alpha \cdot p $}
+\beta MU^{\gamma_5}\right)\psi_\mu
=\epsilon_\mu\psi_\mu \, .
\label{eigen}
\end{eqnarray}
Denoting the corresponding eigenstate $|\mu\rangle$ and the state with
the lowest positive energy eigenvalue $|0\rangle$, the valence quark wave
function $\Psi_{\rm (val)}$ is given as a perturbation expansion in the
angular velocities $(i/2)\mbox {\boldmath $\tau \cdot \Omega$}
=R^\dagger(t) \dot R(t)$:
\begin{eqnarray}
\Psi_{\rm (val)}= R \left(\psi_0+\frac{1}{2}\sum_{\mu\ne0}\psi_\mu
\frac{\langle\mu|\mbox {\boldmath $\tau \cdot \Omega$}|0\rangle}
{\epsilon_0-\epsilon_\mu}\right)+ {\cal O}(\mbox {\boldmath $\Omega$}^2).
\label{psival}
\end{eqnarray}
Substitution of eqn. (\ref{psival}) into eqn. (\ref{aval1}) yields
\begin{eqnarray}
A_i^{{\rm (val)}a} = N_C
D_{ab} \psi_0^\dagger\alpha_i\gamma_5 \frac{\tau^b}{2}\psi_0
&+&\frac{1}{2}\left(\Omega_j D_{ab}\right)
\sum_{\mu\ne0}\frac{\langle\mu|\tau_j|0\rangle}{\epsilon_0-\epsilon_\mu}
\psi_\mu^\dagger\alpha_i\gamma_5 \frac{\tau^b}{2}\psi_0
\nonumber \\
&+&\frac{1}{2}\left(D_{ab} \Omega_j \right)
\sum_{\mu\ne0}\frac{\langle0|\tau_j|\mu\rangle}{\epsilon_0-\epsilon_\mu}
\psi_0^\dagger\alpha_i\gamma_5 \frac{\tau^b}{2}\psi_\mu
+{\cal O}(\mbox {\boldmath $\Omega$}^2).
\label{aval2}
\end{eqnarray}
The terms linear in $\mbox {\boldmath $\Omega$}$ on the RHS of
eqn. (\ref{aval2}) give vanishing contribution to $g_A$ as long as the
angular velocities are considered to be pure c-numbers; {\it i. e.}
there is ${\underline {\rm no}}$ $1/N_C$ correction to $g_A$ if
$\mbox {\boldmath $\Omega$}$ is treated as a classical
quantity. However, this is no longer the case when the semiclassical
quantization prescription
\begin{eqnarray}
&&\Omega_i\longrightarrow \frac{J_i}{\alpha^2}
\label{quant}
\end{eqnarray}
is imposed. Here $J_i$ are the spin operators in the space of the
collective coordinates and $\alpha^2$ denotes the (iso-) rotational
moment of inertia. The analytic expression for $\alpha^2$ in the NJL
model may be found in ref. \cite{re89}. Using furthermore the commutation
relation
\begin{eqnarray}
&&[J_j,D_{ab}]=i\epsilon_{jbm}D_{am}
\label{quanthelp}
\end{eqnarray}
the collective operator involved in the correction to $g_A$ is then
found to be of the same structure as in the leading term in agreement
with more general considerations \cite{da93}. Up to linear order in
$1/\alpha^2$ the valence quark contribution to the nucleon axial
coupling constant is obtained to be
\cite{wa93}:
\begin{eqnarray}
g^{\rm (val)}_A = -\frac{N_C}{3}\left(\langle 0|\Sigma_3\tau_3|0\rangle
+\frac{i}{\alpha^2}\sum_{\mu\ne0}
\frac{\langle 0|\tau_1|\mu\rangle
\langle\mu|\Sigma_3\tau_2|0\rangle}
{\epsilon_0-\epsilon_\mu}\right).
\label{gacorr}
\end{eqnarray}
As $\alpha^2$ is of the order $N_C$ the correction in (\ref{gacorr})
is of order $1/N_C$. The result (\ref{gacorr}) is, of course, ambiguous.
This is due to ordering ambiguities of the semiclassical quantization
prescription(\ref{quant}) arising when the classical collective coordinates
are substituted by the collective generators. Since there is no derivation
of this prescription from first principles uncertainties of this kind are
unavoidable. Ignoring, for the time being, this kind of ambiguities we
will explain the relevance of PCAC once this special quantization prescription
is adopted. In order to derive the corresponding $1/N_C$ corrections to
the equation of motion we need to evaluate the divergence of (\ref{aval2}).
Using the eigenequation (\ref{eigen}) we find:
\begin{eqnarray}
\frac{1}{N_C}\hat r_b D_{ab} \partial_i A_i^{{\rm (val)}a} & = &
M \psi_0^\dagger\beta\left(-{\rm sin}\Theta+
i\mbox{\boldmath $\tau$} \cdot{\bf \hat r}\gamma_5
{\rm cos}\Theta\right)\psi_0
\nonumber \\
&&\hspace{0.2cm} + \frac{iM}{4\alpha^2}\sum_{\mu\ne0}
\frac{\langle0|\tau_j|\mu\rangle}{\epsilon_0-\epsilon_\mu}
\psi_\mu^\dagger\beta\gamma_5
\big[\mbox{\boldmath $\tau$} \cdot{\bf \hat r},\tau_j\big]
\psi_0{\rm cos}\Theta.
\label{eqmcorr}
\end{eqnarray}
The first term on the $RHS$ of eqn. (\ref{eqmcorr}) is just
the valence quarks' contribution to the classical equation of
motion \cite{re88} for chiral angle while the second term
represents the $1/N_C$ correction we are looking for.
According to ref.\  \cite{wa93} no $1/N_C$ corrections to
$g_A$ arise from the sea-part of the action ${\cal A}_{\rm sea}$:
\begin{eqnarray}
g_A^{\rm sea} = \frac{N_C}{6}\sum_{\mu}{\rm erfc}
(|\frac{\epsilon_\mu}{\Lambda}|)
\langle\mu|\Sigma_3\tau_3|\mu\rangle
\label{gasea}
\end{eqnarray}
and thus there is also no further change to equation of motion. Note that
in the chiral limit ($m_\pi=0$) an overall factor 3/2 appears in
eqns. (\ref{gacorr},\ref{gasea}) \cite{ad83}. The absence of $1/N_C$
corrections to $g_A^{\rm sea}$ has been doubted and small corrections
may indeed be present \cite{goepriv}. We will not go into this point
further but rather concentrate on the $1/N_C$ corrections to the
equation of motion stemming from the valence quark wave function
since this suffices to illuminate the relevance of PCAC.
The complete equation of motion we would like to solve then reads
\begin{eqnarray}
&&{\rm cos}\Theta(r)\left[\int d\Omega\left(
\bar\psi_0 i\mbox{\boldmath $\tau$} \cdot{\bf \hat r}\gamma_5 \psi_0
+\frac{i}{4\alpha^2}\sum_{\mu\ne0}
\frac{\langle0|\tau_j|\mu\rangle}{\epsilon_0-\epsilon_\mu}
\bar\psi_\mu\gamma_5 \big[\mbox{\boldmath $\tau$} \cdot{\bf \hat r},
\tau_j\big] \psi_0 +{\rm tr}(i\mbox{\boldmath $\tau$} \cdot{\bf \hat r}
\gamma_5 \rho_S^{\rm sea})\right)\right]
\nonumber \\
&&\hspace{1cm} =
{\rm sin}\Theta(r)\left[\int d\Omega\left( \bar\psi_0 \psi_0
+{\rm tr}(\rho_S^{\rm sea})\right)
-\frac{4\pi}{N_C}\frac{m_\pi^2f_\pi^2}{M}\right]
\label{eqmcomplete}
\end{eqnarray}
wherein $\rho_S^{\rm sea}$ denotes the sea contribution to the
scalar density defined in ref. \cite{re88}.

As the moment of inertia $\alpha^2$ is a functional of the meson profile
the appearance of $\alpha^2$ in (\ref{eqmcomplete})
transforms the original algebraic equation into an integral equation. Its
numerical solution is obtained by iteration. For a given
$\alpha^2$ the self-consistent solution to eqns. (\ref{eigen})
and (\ref{eqmcomplete}) is constructed. This solution then yields
a new $\alpha^2$ to be substituted in eqn. (\ref{eqmcomplete}).
Since the resulting $\alpha^2$ deviates only moderately from the
one obtained when the $1/N_C$ correction is omitted a converging
solution is already obtained after a few iterations.

\begin{figure}
\centerline{
\psfig{figure=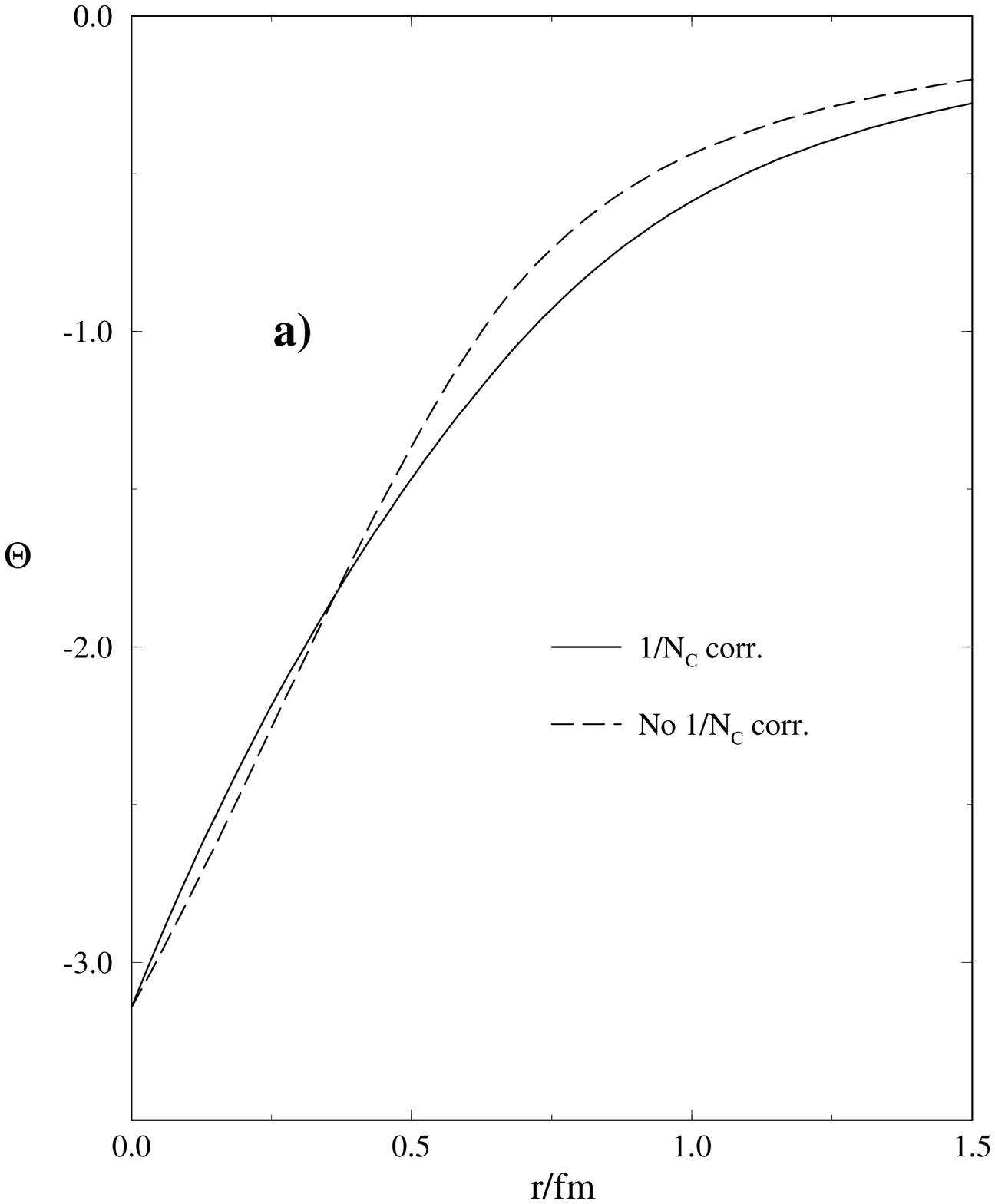,height=11cm,width=9cm}
\hfill
\psfig{figure=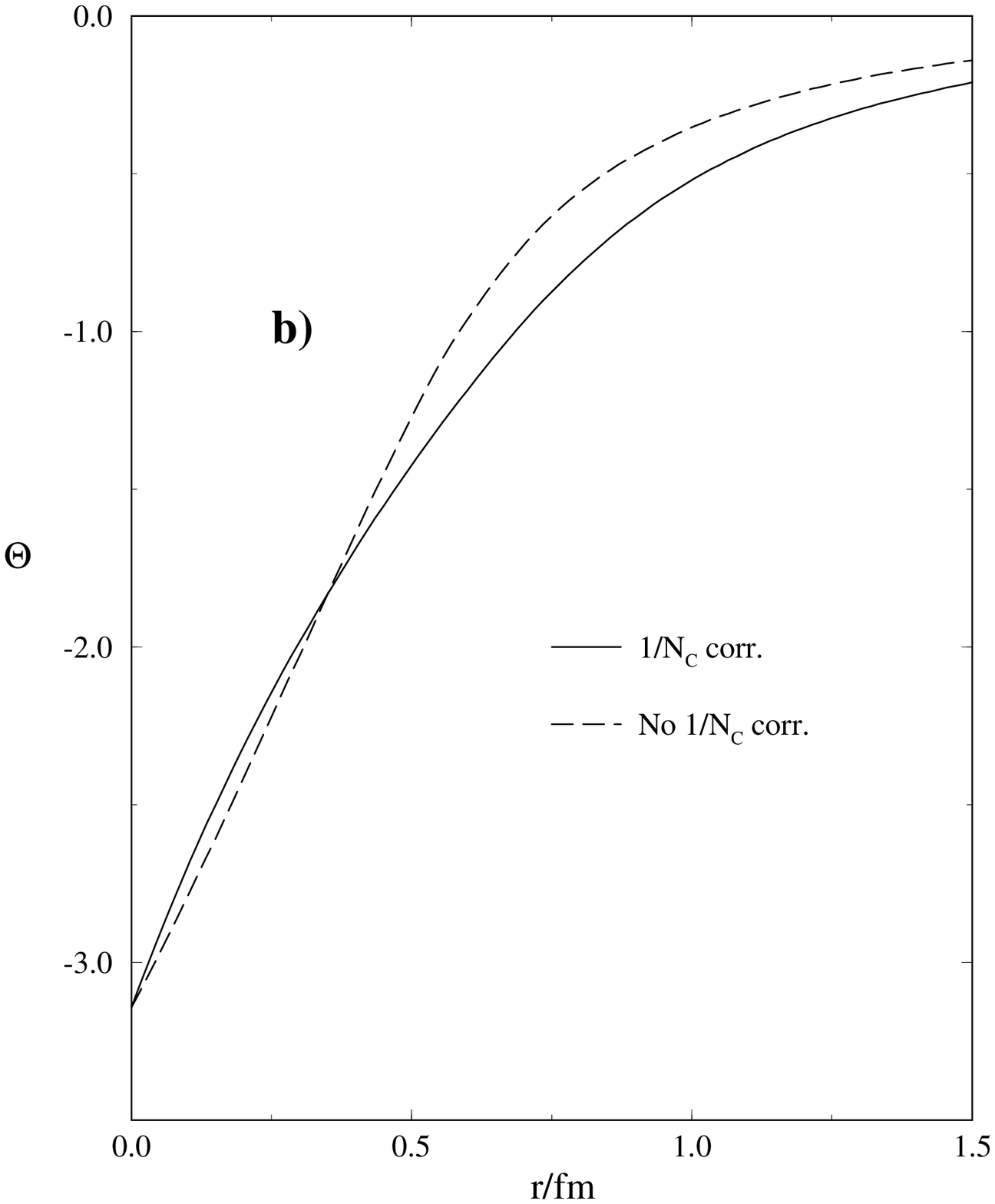,height=11cm,width=9cm}
}
\vspace{1cm}
\fcaption{The chiral angle $\Theta$ as the self-consistent solution of
eqns. (\ref{eigen},\ref{eqmcomplete}): solid lines. The dashed lines refer
to the omission of the $1/\alpha^2$ term in (\ref{eqmcomplete}). Displayed
are the cases $m_\pi=0$ (a) and $m_\pi=138$MeV (b).}
\label{fig1}
\end{figure}

\begin{table}
\tcaption{Static properties for the soliton solution of eqns.
(\ref{eigen},\ref{eqmcomplete}): `$\Theta$ corr'. `$\Theta$ not corr'
refers to the omission of the $1/\alpha^2$ term in the equation of
motion (\ref{eqmcomplete}). The superscript for $g_A$ denotes the order
of $1/\alpha^2$ in eqns. (\ref{gacorr}) and (\ref{gasea}).
$g_A^{\rm (tot)}=g_A^{(0)}+g_A^{(1)}$ is the final result for the
nucleon axial coupling constant. $\epsilon_0$, $E_{total}$,
$\alpha^2$ and $\langle r^2\rangle _{I=0}^{1/2}$ are the valence
quark eigenenergy, total energy of the soliton, moment of inertia and
the isoscalar root mean square radius, respectively. We have used the
constituent quark mass $M=400$MeV.}
\newline
\centerline{\tenrm\smalllineskip
\begin{tabular}{|l|c|c|c|c|}
\hline
& \multicolumn{2}{c|} {$m_\pi=0$} &
 \multicolumn{2}{c|} {$m_\pi=138$MeV}   \\
\hline
&$\Theta$ not corr.&$\Theta$ corr.&
$\Theta$ not corr.&$\Theta$ corr.\\
\hline
$g_A^{(0)}$ & 0.83 & 0.82 & 0.75 & 0.81 \\
\hline
$g_A^{(1)}$ & 0.47 & 0.35 & 0.39 & 0.30 \\
\hline
$g_A^{\rm (tot)}$ & 1.30 & 1.17 & 1.14 & 1.12 \\
\hline
$g_A$ from eqn. (\ref{tail}) & 0.82 & 1.13 & 0.76 & 1.14 \\
\hline
$\epsilon_0$/MeV & 190 & 151 & 215 & 167 \\
\hline
$E_{total}$/MeV & 1203 & 1233 & 1253 & 1266\\
\hline
$\alpha^2$/(1/GeV) & 6.49 & 7.50 & 5.86 & 6.09 \\
\hline
$\langle r^2\rangle _{I=0}^{1/2}$/fm&  0.90  & 0.98 &
0.75& 0.80 \\
\hline
\end{tabular}}
\label{tab1}
\end{table}

As can be seen from fig.\ (1) the pion profile at large radii
is increased if the $1/N_C$ correction is taken into account. This
increases the moment of inertia $\alpha ^2$ which in turn decreases
the correction term in the equation of motion (\ref{eqmcomplete}) and
in the expression for $g_A$ (\ref{gacorr}). This behavior guarantees the
stability of the $1/N_C$ corrected soliton and limits the change of most
quantities, see table (1). {\it E.g.} the soliton energy $E_{total}$
\cite{re89} is increased by a few percent only. As the slope of the soliton
profile close to the origin becomes moderately larger, the valence quark
energy $\epsilon_0$ decreases accordingly. Also the moment of inertia
increases by a small amount yielding an even smaller nucleon--$\Delta$
mass splitting $M_\Delta-M_n=3/2\alpha^2$. A similar behavior is found
for the isoscalar root mean square radius
$\langle r^2\rangle _{I=0}^{1/2}$. In the chiral limit it is even
overestimated when the $1/N_C$ correction is taken into account. In
contrast to the moderate change of other quantities the first order
correction for $g_A$ drops significantly. Whereas in the chiral limit
this change is reflected also in the total sum for $g_A$ there is almost
no net effect in the case of the physical pion mass because the zeroth
order value for $g_A$ increases.

Table (1) reveals also that without correction for the meson profile the
value of $g_A$ differs significantly from the value as obtained using
eq.\ (\ref{tail}), {\it i.e.} PCAC is strongly violated. As expected
the PCAC relation (\ref{tail}) provides values which are equal to the
zeroth order contribution. On the other hand, with the corrected meson
profile PCAC is fulfilled within numerical uncertainties. Note that
the extraction of $g_A$ from (\ref{tail}) becomes the more difficult
the larger the soliton is, as all numerical  calculations are done in
finite boxes.
We would also like to mention that we find some discrepancies from the
numerical results of ref. \cite{wa93}. Some of these discrepancies can be
resolved by multiplying their result for $g_A^{(1)}$ for $m_\pi=0$ by 3/2.
The necessity of this factor has been pointed out in ref. \cite{ad83}.

In conclusion, we have calculated the NJL soliton including $1/N_C$
corrections as they result from the ambiguous quantization ``recipe''
(\ref{quant}). As claimed already in ref.\ \cite{wa93} these
corrections solve the problem of a too small value of the axial
coupling of the nucleon $g_A$ in hedgehog soliton models.  Taking into
account these corrections in a way that PCAC is still fulfilled decreases
the value of the correction for $g_A$. Nevertheless, the total result is
less than ten percent off the experimental value. Here we argued on
the basis of PCAC that the soliton profile has to change also if we
adopt the quantization (\ref{quant},\ref{quanthelp}). It would be
interesting to construct the energy functional which corresponds to
that operator ordering and therefore provides the meson profile
calculated in this letter from a variational principle. Since the
soliton mass is of order $N_C$ this correction has to be of order
$N_C^0$. On the other hand, as long as the existence of such a
formalism cannot be shown, the application of ``recipe''
(\ref{quant},\ref{quanthelp}) to different orderings of collective
coordinates and operators will at least be doubtful. After all,
it would be somewhat astonishing if the problem of a too small $g_A$
for hedgehog soliton models turned out to be only a result of a peculiar
quantization.

\vspace{3cm}
\nonumsection{\large \bf Acknowledgement}
We are grateful to H. Reinhardt for stimulating
discussions and helpful comments on the manuscript.

\vfil\eject

\vfil\eject

\pagestyle{empty}

\centerline{\Large \bf Table caption}

\vspace{1cm}

\noindent
Table 1: Static properties for the soliton solution of eqns.
(\ref{eigen},\ref{eqmcomplete}): `$\Theta$ corr'. `$\Theta$ not corr'
refers to the omission of the $1/\alpha^2$ term in the equation of
motion (\ref{eqmcomplete}). The superscript for $g_A$ denotes the order
of $1/\alpha^2$ in eqns. (\ref{gacorr}) and (\ref{gasea}).
$g_A^{\rm (tot)}=g_A^{(0)}+g_A^{(1)}$ is the final result for the
nucleon axial coupling constant. $\epsilon_0$, $E_{total}$,
$\alpha^2$ and $\langle r^2\rangle _{I=0}^{1/2}$ are the valence
quark eigenenergy, total energy of the soliton, moment of inertia and
the isoscalar root mean square radius, respectively. We have used the
constituent quark mass $M=400$MeV.

\vspace{3cm}
\centerline{\tenrm\smalllineskip
\begin{tabular}{|l|c|c|c|c|}
\hline
& \multicolumn{2}{c|} {$m_\pi=0$} &
 \multicolumn{2}{c|} {$m_\pi=138$MeV}   \\
\hline
&$\Theta$ not corr.&$\Theta$ corr.&
$\Theta$ not corr.&$\Theta$ corr.\\
\hline
$g_A^{(0)}$ & 0.83 & 0.82 & 0.75 & 0.81 \\
\hline
$g_A^{(1)}$ & 0.47 & 0.35 & 0.39 & 0.30 \\
\hline
$g_A^{\rm (tot)}$ & 1.30 & 1.17 & 1.14 & 1.12 \\
\hline
$g_A$ from eqn. (\ref{tail}) & 0.82 & 1.13 & 0.76 & 1.14 \\
\hline
$\epsilon_0$/MeV & 190 & 151 & 215 & 167 \\
\hline
$E_{total}$/MeV & 1203 & 1233 & 1253 & 1266\\
\hline
$\alpha^2$/(1/GeV) & 6.49 & 7.50 & 5.86 & 6.09 \\
\hline
$\langle r^2\rangle _{I=0}^{1/2}a$/fm&  0.90  & 0.98 &
0.75& 0.80 \\
\hline
\end{tabular}}

\vspace{4cm}

\centerline{\Large \bf Figure caption}

\vspace{1cm}

\noindent
Figure 1: The chiral angle $\Theta$ as the self-consistent solution of
eqns. (\ref{eigen},\ref{eqmcomplete}): solid lines. The dashed lines refer
to the omission of the $1/\alpha^2$ term in (\ref{eqmcomplete}). Displayed
are the cases $m_\pi=0$ (a) and $m_\pi=138$MeV (b).

\eject

\begin{figure}
\vskip1cm
\centerline{\hskip -1.5cm
\psfig{figure=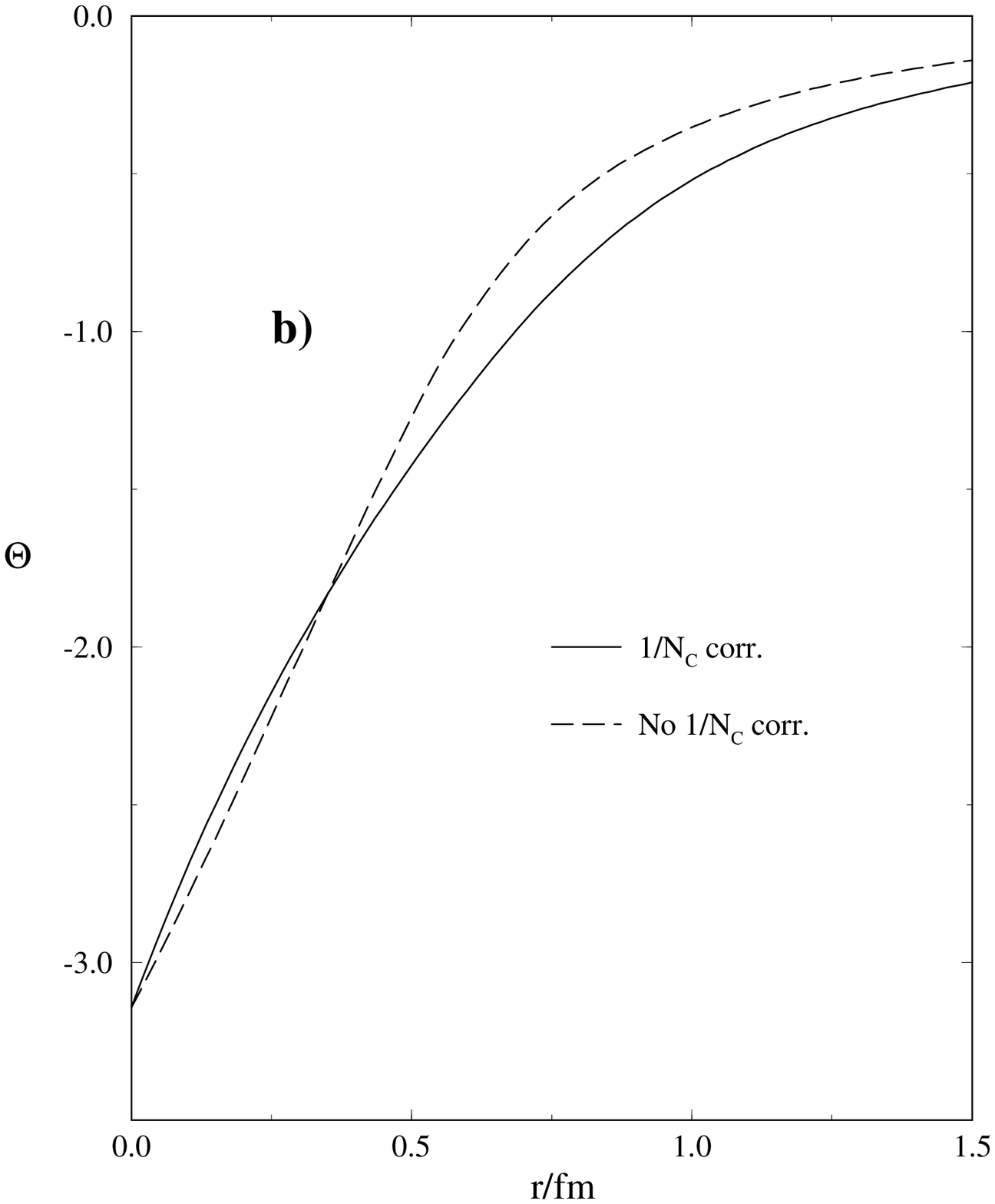,height=11.0cm,width=16cm}}
\vskip2.0cm

\centerline{\hskip -1.5cm
\psfig{figure=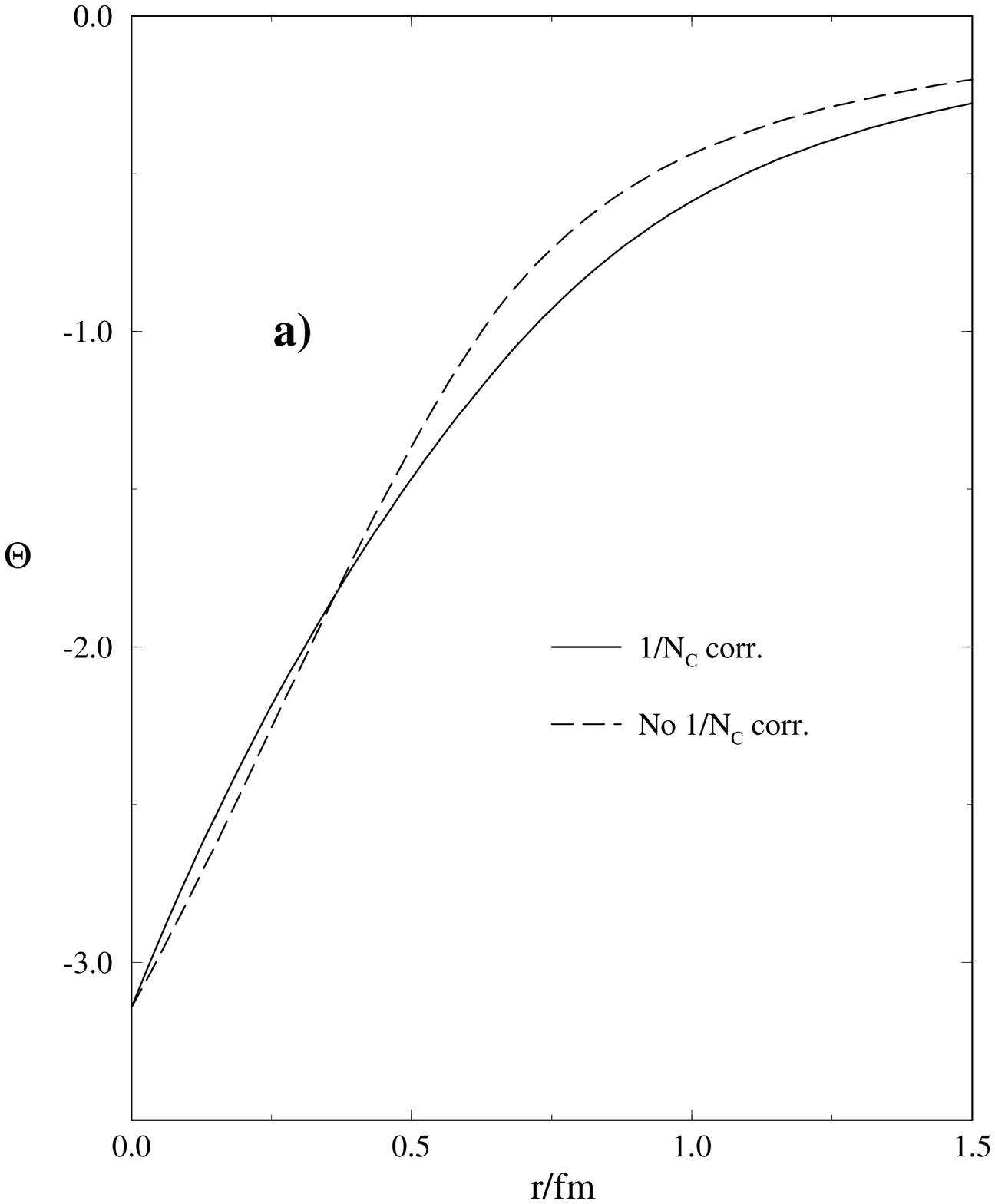,height=11.0cm,width=16cm}}
\end{figure}

\end{document}